\documentclass[useAMS,usenatbib]{mn2e}
\usepackage{graphicx,amssym,color}
\newcommand{\be}{\begin{equation}}
\newcommand{\ee}{\end{equation}}

\def\ltsima{$\; \buildrel < \over \sim \;$}
\def\simlt{\lower.5ex\hbox{\ltsima}}
\def\gtsima{$\; \buildrel > \over \sim \;$}
\def\simgt{\lower.5ex\hbox{\gtsima}}
\def\la{$\langle$}

\title{The Dynamical State and Mass-Concentration Relation of Galaxy Clusters}

\author[Ludlow et al.] {\parbox{18cm}{
Aaron D. Ludlow$^{1,\star}$,
Julio F. Navarro$^{2}$,
Ming Li$^{3}$,
Raul E. Angulo$^{3}$,
Michael Boylan-Kolchin$^{4}$,
and Philip E. Bett$^{1}$,
}\vspace{0.3cm}\\
$^{1}${Argelander-Institut f\"{u}r Astronomie, Auf dem H\"{u}gel 71,
D-53121 Bonn, Germany}\\
$^{2}${Dept. of Physics and Astronomy, University of
    Victoria, Victoria, BC, V8P 5C2, Canada}\\
$^{3}$Max-Planck-Institut f\"{u}r Astrophysik,
Karl-Schwarzschild-Stra\ss{}e 1, 85740 Garching bei M\"{u}nchen,
Germany\\
$^{4}${Center for Galaxy Evolution, 4129 Reines Hall, University of
  California, Irvine, CA 92697, USA}\\
}

\begin{document}
\maketitle 

\begin{abstract}
  We use the Millennium Simulation series to study how the dynamical
  state of dark matter halos affects the relation between mass and
  concentration. We find that a large fraction of massive systems are
  identified when they are substantially out of equilibrium and in a
  particular phase of their dynamical evolution: the more massive the
  halo, the more likely it is found at a transient stage of high
  concentration. This state reflects the recent assembly of massive
  halos and corresponds to the first pericentric passage of
  recently-accreted material when, before virialization, the kinetic
  and potential energies reach maximum and minimum values,
  respectively. This result explains the puzzling upturn in the
  mass-concentration relation reported in recent work for massive
  halos; indeed, the upturn disappears when only dynamically-relaxed
  systems are considered in the analysis. Our results warn against
  applying simple equilibrium models to describe the structure of rare,
  massive galaxy clusters and urges caution when extrapolating scaling
  laws calibrated on lower-mass systems, where such deviations from
  equilibrium are less common.
  The evolving dynamical state of galaxy
  clusters ought to be carefully taken into account if cluster studies
  are to provide precise cosmological constraints.
\end{abstract}

\renewcommand{\thefootnote}{\fnsymbol{footnote}}
\footnotetext[1]{E-mail:aludlow@astro.uni-bonn.de} 

\section{Introduction}
\label{sec:intro}

Galaxy clusters are powerful cosmological probes. Because of their
large masses, they can be detected at cosmological distances despite
being rare and studied at various wavelengths, from X-rays to optical to
millimeter wavelengths. Their rarity is actually a strength when
it comes to cosmology; massive clusters trace the tail of high-mass
objects able to collapse under their own gravity and therefore their
numbers, as a function of mass and redshift, are exponentially
sensitive to the normalization of the power spectrum of density
fluctuations and to the cosmological parameters that govern the
universal expansion history \citep[see, e.g.,][for a recent review and
a complete list of references]{Allen2011}. 

Cluster counts are thus widely recognized as a premier tool able to
provide cosmological constraints complementary to those based on
analysis of the cosmic microwave background, supernova luminosity
distances, galaxy clustering, and gravitational lensing \citep[see,
e.g.,][]{Mantz2008,Cunha2009,Henry2009,Rozo2009,Vikhlinin2009}.  Their
potential is more directly realized, however, when observables can be
turned into effective measures of cluster mass; indeed, the most
robust and discriminating theoretical predictions concern the
abundance and redshift evolution of clusters of different mass.

Mass, however, is not directly observable, which means in practice
that observables such as cluster richness, velocity dispersion, X-ray
luminosity/temperature, or Sunyaev-Zel'dovich (SZ) decrement must be
related to mass through scaling laws whose shape and scatter need
careful calibration. Although virial equilibrium and self-similarity
suggest power-law scalings between mass and observational
proxies \citep{Kaiser1986,Navarro1995}, it has long been appreciated
that baryon physics breaks self-similarity and imposes scalings with
different slopes and redshift dependence \citep[e.g.,][]{Kaiser1991,Evrard1991}.

The complexity of the baryon physics involved precludes robust theoretical
predictions and therefore the mass-observable relations are usually
parameterized as power laws of adjustable slope, redshift-dependent
normalization, and Gaussian scatter that are empirically calibrated
using a (usually small) set of well-studied clusters. Besides
simplicity, the underlying rationale of this procedure is that, however intricate, the
mass-dependence of the physical processes responsible for a given
mass-observable relation is monotonic and that deviations of
individual clusters from the mean trends are driven by stochastic
effects.

The validity of these assumptions, however, is not assured and recent
work has highlighted their limitations. One example is the mismatch
between the average SZ signal at given optical richness measured by
the Planck satellite and expectations based on the X-ray properties of
clusters with weak lensing-calibrated masses \citep{Planck2011c}. This
might either indicate a dichotomy in the gas content of
clusters of different mass or perhaps arises from the large, correlated
scatter between observables suggested by recent simulation work
\citep[see, e.g.,][]{Angulo2012}.

A second example is provided by recent reports of a non-monotonic
relation between the mass and concentration of massive dark matter
halos \citep{Klypin2011,Prada2011}. These authors report an ``upturn''
in the {\it median} concentration on mass scales where the rms
fluctuations of the linear density field, $\sigma(M,z) \simlt
1/2$. This corresponds to rare objects with virial masses equivalent
to a few times $10^{15}\, M_\odot$ at $z=0$, shifting to lower masses
at higher redshift.

The origin of the upturn is unclear. Since halo concentration reflects
the mean density of the universe at the time of assembly
\citep{Navarro1996,Navarro1997,Eke2001,Bullock2001,Wechsler2002,Zhao2003a},
the mass-concentration relation is expected to be a monotonic one
where concentration decreases with increasing halo mass. An upturn in
concentration may have non-negligible consequences on cluster studies:
concentration is a measure of the characteristic density of a cluster,
and it could therefore affect directly mass proxies that are sensitive
to density, such as X-ray luminosity. A non-monotonic
mass-concentration relation might thus lead to selection biases or
``breaks'' in the relation between X-ray properties and mass that
would be poorly captured by the assumed power-law scalings.

We use here data from the Millennium Simulation series to revisit the
mass-concentration relation of rare, massive halos. In particular, we
examine the physical origin of the reported ``upturn'' in
concentration and the nature of the scatter about the mean
relation. We describe briefly the numerical simulations in
Sec.~\ref{SecNumSims}, present our results in Sec.~\ref{SecRes}, and
summarize our main conclusions in Sec.~\ref{SecConc}.

\section{Numerical Simulations}
\label{SecNumSims}

\subsection{The Millennium Simulation series}
\label{SecMillSims}

We use in our analysis halos identified in the three ``Millennium
Simulations'', which we refer to as MS-I, MS-II, and MS-XXL, or
collectively as MS.  All MS runs adopted the same cosmology (a WMAP-1
normalized flat $\Lambda$CDM model) and the same sequence of outputs
times in order to facilitate comparisons between the runs. The values
of the cosmological parameters are as follows: $\Omega_{\rm m}=0.25$,
$\Omega_{\Lambda}=1-\Omega_{\rm m}=0.75$, $h=0.73$, $n=1$ and
$\sigma_8=0.9$. Here $\Omega_i$ is the present-day contribution of
component $i$ to the universal matter energy density in units of the
critical density for closure; $\sigma_8$ is the rms mass fluctuation
in spheres of $8 \, h^{-1}$ Mpc radius linearly extrapolated to $z=0$;
$n$ is the spectral index of primordial density fluctuations, and $h$
is the Hubble parameter defined so that $H_0=H(z=0)=100 \, h$
km/s/Mpc.

The MS-I used $2160^3$ particles of mass $m_p=8.61\times 10^8 \ h^{-1}
M_{\odot}$ to follow the evolution of the dark matter component in a
periodic box $500 \ h^{-1} {\rm Mpc}$ on a side
\citep[]{Springel2005a}. MS-II used the same number of particles, but
a box size $5$ times smaller, $L_{\rm box}=100 \ h^{-1} {\rm Mpc}$
\citep[]{Boylan-Kolchin2009}; each MS-I particle is thus $125\times$
heavier than in MS-II. The MS-XXL is substantially larger than MS-I,
both in box size ($L_{\rm box}=3 \, h^{-1} {\rm Gpc}$) and total
particle number ($6720^3$), and thus provides the best statistics for
the rarest and most massive objects at any redshift
\citep[]{Angulo2012}, albeit with poorer mass resolution. The particle
mass in the XXL is $m_p=6.17\times10^9 \, h^{-1} M_{\odot}$. The
Plummer-equivalent gravitational softening lengths used in these runs
are $\epsilon_P=5 \, h^{-1} {\rm kpc}$ (MS-I), $1 \, h^{-1} {\rm kpc}$
(MS-II), and $10 \, h^{-1} {\rm kpc}$ (MS-XXL).


\begin{figure*}
\begin{center}
\resizebox{13cm}{!}{\includegraphics{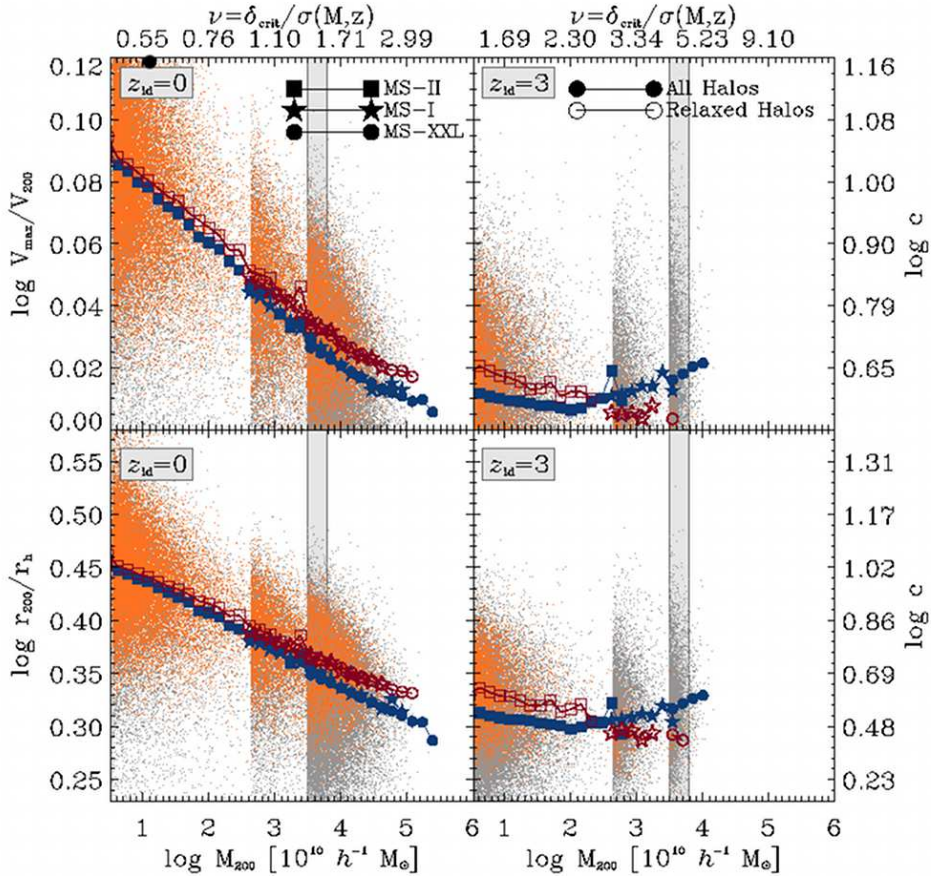}}
\end{center}
\caption{Mass-concentration relation for all halos resolved with more
  than $5000$ particles in the Millennium Simulations. The top panels
  show, as a function of virial mass, concentrations estimated via
  $V_{\rm max}/V_{200}$, the ratio between the maximum circular
  velocity and the virial velocity of a halo. Bottom panels show
  concentrations estimated by the ratio between the half-mass radius
  and virial radius, $r_{200}/r_{\rm h}$. For each set of panels, the
  corresponding NFW concentration values are indicated by the
  right-hand side tickmarks. Panels on the left correspond to halos
  identified at $z_{\rm id}=0$, those on the right at $z_{\rm
    id}=3$. The three groups of points in each panel correspond, from
  left to right, to MS-II, MS-I and MS-XXL halos, respectively. All
  halos are shown in grey, relaxed halos in orange. Connected symbols
  (see legends) indicate the median concentration computed in bins of
  virial mass. Open and filled symbols correspond to ``relaxed'' or
  all halos, respectively.  Note the excellent agreement in the median
  concentration of halos of similar mass identified in different
  simulations. The bottom tickmarks indicate halo virial mass; those
  at the top the dimensionless mass measure,
  $\nu=\delta_{\rm crit}/\sigma(M,z)$. Note that, because of the fixed comoving
  box size of each run, halos identified at $z=3$ sample ``rarer'' peaks in the
  mass spectrum (i.e., higher values of $\nu$) than those at
  $z=0$.}
\label{FigMc}
\end{figure*}

\subsection{Analysis}
\label{SecAnal}

Halos in the MS are identified using a friends-of-friends (FOF) group
finder \citep{Davis1985} with a linking length equal to 20\% of the
mean interparticle separation. Substructure in each FOF group is
identified using {\small SUBFIND} \citep{Springel2001b}. We consider
in our analysis only the main halo of each FOF grouping, and define
its virial mass, $M_{200}$, as the mass contained within a sphere
centered at the potential minimum that encloses a mean overdensity of
$200$ times the critical density for closure, $\rho_{\rm
  crit}=3H(z)^2/8\pi G$.  This implicitly defines the virial radius of
a halo, $r_{200}$, and its virial velocity, $V_{200}=\sqrt{G M_{200} /
  r_{200}}$.

We adopt two different quantities to estimate halo concentrations,
$c$. One uses the ratio of the peak circular speed, $V_{\max}$, to the
virial velocity, $V_{200}$, as in \citet{Prada2011}. Concentrations
are derived from $V_{\max}/V_{200}$ assuming that the mass profile
follows the NFW formula \citep{Navarro1996,Navarro1997}. We use {\it
  all} particles within $r_{200}$ to estimate $c$; i.e., $V_{\rm max}$
is computed without subtracting substructure. This is important in
order to ensure convergence on the overlapping mass scales of
different MS runs because substructure, which may affect
the estimates, is particularly sensitive to numerical
resolution\footnote{The slight concentration mismatch betwen MS-I and
  MS-II halos reported by \citet{Prada2011} is in all likelihood due
  to this effect (see Fig.~\ref{FigMc}).}.

Some advantages of this estimator include the fact that it is simple
to compute, fairly stable numerically ($V_{\rm max}$ is well defined
and robustly estimated even at modest numerical resolution), and that
it does not rely on profile fitting and residual-minimization
techniques. The main disadvantage is that, because it assumes an NFW
profile, it is unduly sensitive when $V_{\rm max}$ approaches
$V_{200}$: for example, when concentrations double from $c=2.5$ to $5$
$V_{\rm max}/V_{200}$ changes only from $1.001$ and $1.062$,
respectively. Even tiny changes in $V_{\rm max}$ may thus lead to
large variations in the derived concentration. Further, a large value
of $V_{\rm max}/V_{200}$ does not necessarily imply a high central
concentration of matter. For example, a massive substructure may
elevate $V_{\rm max}$ but, if located at the outskirts of the halo, it
might actually {\it decrease} the average central density of a halo.

Because of this, we adopt an additional concentration estimator: the
virial-to-half mass radius ratio, $r_{200}/r_{\rm h}$, provides a more reliable
tracer of the concentration of dark matter. As for the velocity ratio,
we use {\it all} particles within $r_{200}$ to compute $r_{\rm h}$ and
express the ratio as a concentration, $c$, assuming an NFW profile.

Our analysis considers all halos resolved with at least 5000 particles
(i.e., $N_{200}\geq 5000$) at three different redshifts, $0$, $1$, and
$3$. This corresponds to a minimum mass of $3.44\times 10^{10} \
h^{-1} M_{\odot}$ for MS-II, $4.30\times 10^{12} \ h^{-1} M_{\odot}$
for MS-I, and $3.09\times 10^{13} h^{-1} M_{\odot}$ for MS-XXL. There
is a fairly large overlap in the mass scales covered by the different
simulations, which allows us to check the sensitivity of our results
to numerical resolution.

As discussed by \citet{Neto2007}, halo concentrations can be affected
by transient departures from equilibrium, {\textbf{resulting in subtle
biases in their mean mass dependence and scatter.}} We therefore identify a
subsample of ``relaxed'' halos, as those that satisfy the following
three criteria: (i) $f_{\rm sub}<0.1$, (ii) $d_{\rm
  off}<0.07$ and (iii) $2T/|U|<1.3$. Here $f_{\rm sub}=M_{\rm
  sub}/M_{200}$ is the mass fraction contributed by substructure;
$d_{\rm off}=|\mathbf{r}_p-\mathbf{r}_{\rm CM}|/r_{200}$ is the offset
between the position of the potential minimum and the halo barycenter,
expressed in units of the virial radius; and $2T/|U|$ is the virial
ratio of kinetic to potential energies. 

A full discussion of these criteria is provided by
\citet{Neto2007}, who point out the need for multiple objective
criteria to reliably assess the equilibrium state of a halo. This is
because all of these measures fluctuate during the virialization
process and may therefore briefly fail to identify
out-of-equilibrium systems. However, the fluctuations of the three
different measures are not synchronized and, therefore, it is
unlikely that all three will fail at the same time. For example, a
halo may have a large abundance of substructure but a small $d_{\rm
  off}$ if its subhalos happen to be distributed approximately
isotropically about the halo center. Conversely, $f_{\rm sub}$ may
be small and $d_{\rm off}$ large when deviations from equilibrium
are driven by a relatively small fraction of the mass, such as in a
succession of minor mergers. Finally, $d_{\rm off}$ and $f_{\rm
  sub}$ make no use of dynamical information and therefore they
often fail at weeding out particular phases of ongoing mergers, for
example, when the merging partner is at pericentric passage. This
configuration minimizes $d_{\rm off}$ (most of the mass is near the
center) and also $f_{\rm sub}$ (substructures are more difficult to
identify when they are close to the center), but their transient
nature would lead to large values of the virial ratio,
$2T/|U|$. The total number of halos considered in each simulation,
as well as the total number of relaxed halos, are listed for each
redshift in Table~\ref{TabHaloSample}.

\begin{figure}
\begin{center}
\resizebox{7.5cm}{!}{\includegraphics{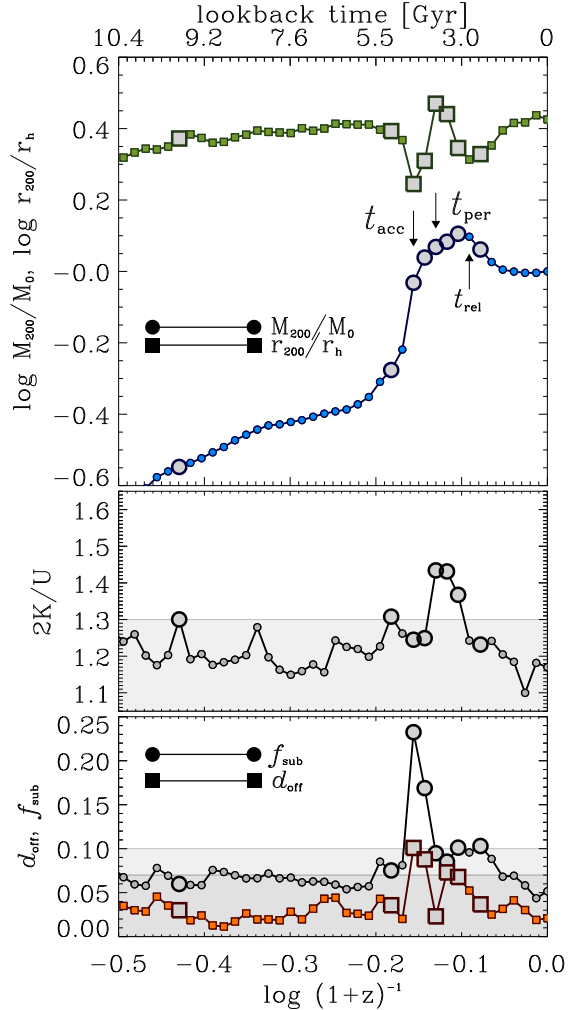}}
\end{center}
\caption{An illustrative example of how concentrations vary during the
  virialization process that follows a major episode of accretion. The
  upper panel shows, as a function of redshift, the evolution of the
  virial mass, $M_{200}$, and of the half-mass radius concentration
  estimator, $r_{200}/r_{\rm h}$, for Aquarius halo F
  \citep{Springel2008b}. This system, which has a half-mass formation
  redshift of $z_{\rm h}=0.56$, is in equilibrium at $z=0$ according
  to all relaxation criteria, whose evolution is plotted in the bottom
  two panels. Large symbols are used to indicate times when the halo
  would not be considered relaxed, according to at least one of the
  relaxation criteria. Although the halo undergoes a major merger at
  $z\sim 0.43$, by $z=0$ more than a crossing time has elapsed,
  allowing the system to reach equilibrium. During virialization,
  however, the concentration estimator, $r_{200}/r_{\rm h}$,
  fluctuates by almost a factor of two. It first reaches a minimum
  (corresponding to $c\approx 1.6$, assuming an NFW profile) at
  $t_{\rm acc}$, when the secondary halo first enters the virial
  radius of the main progenitor, quickly followed by a maximum
  ($c\approx 12.0$) at $t_{\rm per}$, when the merging subhalo is at
  first pericentric approach. A second minimum follows next as the
  accreted material reaches apocenter before relaxing to
  equilibrium. Recent accretion events can clearly bias concentration 
  estimates unless steps are taken to ensure that halos are relaxed.}
\label{FigCVir}
\end{figure}

\begin{figure*}
\begin{center}
\resizebox{18cm}{!}{\includegraphics{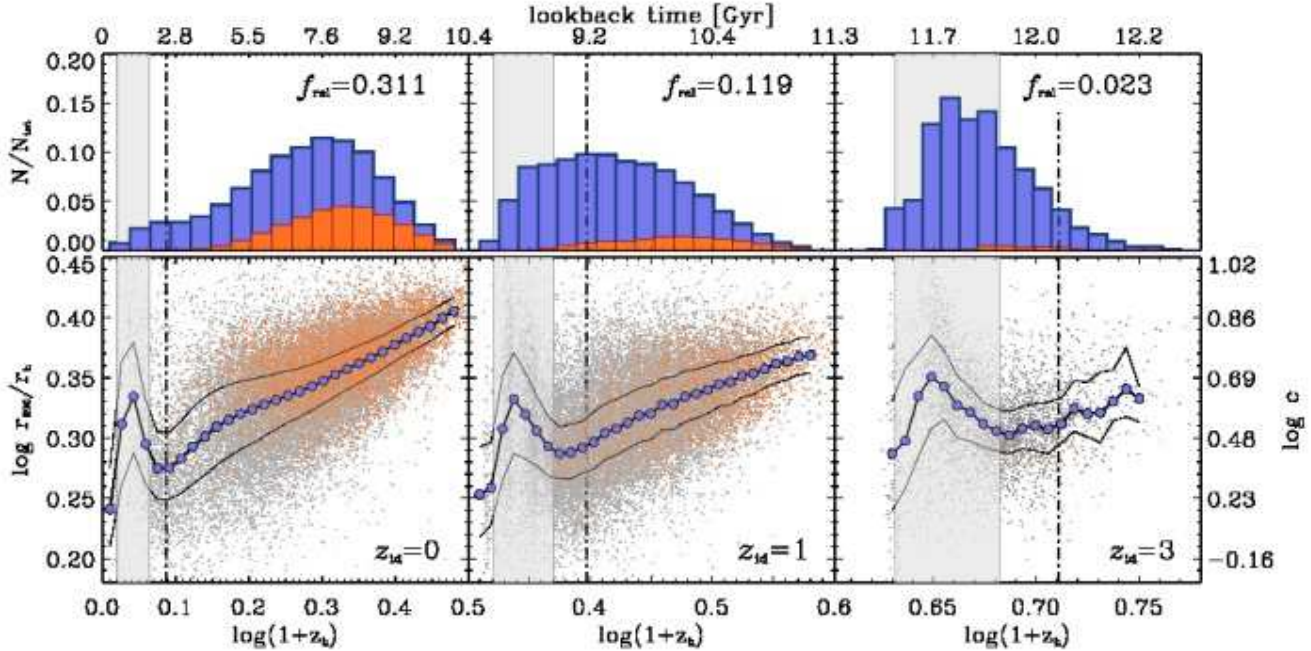}}
\end{center}
\caption{{\it Bottom panels:} Halo concentration, as measured by the
  virial-to-half mass radius ratio, $r_{200}/r_{\rm h}$, versus
  half-mass formation redshift, $z_{\rm h}$, for halos in the mass
  range $3.49 < \log M_{200}/10^{10} h^{-1} M_{\odot} < 3.79$ and
  resolved with at least 5000 particles in all Millennium Simulations
  (see shaded area in Fig.~\ref{FigMc}).  Tickmarks on the right
  indicate concentration values derived assuming that halos follow NFW
  profiles. The different panels show halos identified at three
  different redshifts: $z_{\rm id}=0$, $1$, and $3$.  Grey dots
  indicate all halos; orange dots correspond to halos that pass the
  relaxation criteria specified in Sec.~\ref{SecAnal}. The median
  concentration of the full halo sample is shown by the blue connected
  circles; thin lines delineate the 25th and 75th percentiles. The
  shaded area indicate halos whose formation lookback time since
  identification, $t_{\rm h}=t_{\rm lb}(z_{\rm h})-t_{\rm lb}(z_{\rm
    id})$, is between one quarter and three-quarter crossing times,
  defined as $t_{\rm cross}=2\, r_{200}/V_{200}$. Halos ``formed''
  more than one crossing time ago (to the right of the dot-dashed
  vertical line) are in general close to virial equilibrium; for
  those halos earlier collapse translates into higher
  concentration. The dependence of concentration on formation redshift
  becomes non-monotonic when halos that formed less than one crossing
  time ago are considered, signalling different out-of-equilibrium
  stages in the virialization process. See text for discussion. {\it
    Top panels:} The distribution of formation redshifts for all halos
  shown in the bottom panels (in blue) and for those satisfying the
  relaxation criteria (orange). The fraction of relaxed halos in this
  mass range, $f_{\rm rel}$, decreases with increasing
  redshift. Very few halos formed less than one crossing time ago pass
  the relaxation criteria.}
\label{Figczh}
\end{figure*}

\section{Results}
\label{SecRes}

\subsection{Mass-concentration relation}
\label{ssec:cm_relation}

Figure~\ref{FigMc} shows, as a function of virial mass, the
concentration of halos in our sample, as measured by $V_{\rm
  max}/V_{200}$ (top panels) or $r_{200}/r_{\rm h}$ (bottom
panels). Tickmarks on the right express these quantities in terms of
NFW-based concentration values. All individual halos are shown by grey
dots; those that pass the three relaxation criteria are shown in
orange. Panels on the left show halos identified at $z_{\rm id}=0$;
those on the right correspond to $z_{\rm id}=3$. Median concentrations
for all (blue) and relaxed (red) halos are shown by connected symbols
(see legend on figure for identification). The excellent agreement
between the different runs on overlapping mass scales indicates that
our results are unaffected by numerical resolution effects.

The results for {\it all} halos shown in Fig.~\ref{FigMc} confirm the
upturn in the median concentration of massive halos reported by
\citet{Klypin2011} and \citet{Prada2011}. As discussed by these
authors, the upturn affects only rare, extremely massive systems, and
is therefore more easily appreciated at earlier times in simulations
of fixed comoving box size (i.e., $z_{\rm id}=3$ in Fig.~\ref{FigMc}),
when massive halos trace rarer peaks of the density fluctuation
field. We quantify this using the dimensionless ``peak height''
mass parameter
\begin{equation}
\nu(M,z)=\delta_{\rm crit}(z)/\sigma(M,z),
\label{EqNu}
\end{equation}
defined as the ratio between the critical overdensity for collapse at
redshift $z$ and the linear rms fluctuation at $z$ within spheres
containing mass $M$. The larger the value of $\nu$, the rarer the halo
and the more massive it is relative to the characteristic clustering
mass ($M_*$) at that epoch, which is usually defined by the condition
$\nu(M_*)=1$. Tickmarks along the top of Fig.~\ref{FigMc} list the
values of $\nu$ corresponding to the virial masses listed along the
bottom axes. At $z=0$ even the MS-XXL can only probe the regime
$\nu<3$, whereas at $z=3$ massive halos trace peaks as rare as $\nu\sim
5$.  These objects are considerably rarer than those probed by
\citet{Prada2011} ($\nu\simlt 4$), reflecting the extremely large
volume of MS-XXL.

Interestingly, the upturn disappears when only ``relaxed'' halos are
considered, a clear indication that out-of-equilibrium effects are
responsible. This seems at odds with \citet{Prada2011}, who argue that
the upturn remains even when unrelaxed halos are excluded; however,
they use less restrictive relaxation criteria than the ones we adopt
here. In particular, Prada et al use $d_{\rm off}<0.1$ and
$2T/|U|<1.5$, compared with our adopted values of $0.07$ and $1.3$,
respectively, which are based on the work of \citet{Neto2007}. These
relatively small differences in the criteria have a large effect on
the massive halo sample.  For example, at $z=3$ fewer than $\sim 3\%$
of MS-XXL halos in the sample satisfy our offset and virial ratio
criteria; on the other hand, $61\%$ of them pass the less restrictive
criteria adopted by Prada et al.

If the upturn is driven by unrelaxed halos then we should be able to
get some clues to its origin by considering how concentration
estimates vary as halos virialize.  We turn our attention to this
point next.

\begin{figure*}
\begin{center}
\resizebox{18cm}{!}{\includegraphics{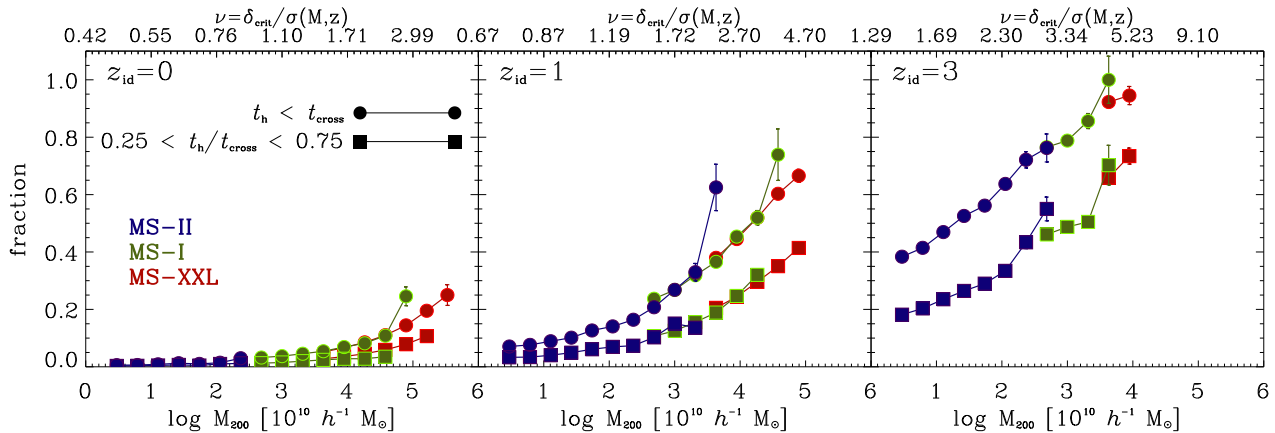}}
\end{center}
\caption{Fraction of halos identified at three different redshifts in
  different stages of relaxation shown as a function of virial
  mass. The top curves correspond to halos that ``formed'' less than one
  crossing time ago; i.e., those where $t_{\rm h}=t_{\rm lb}(z_{\rm
    h})- t_{\rm lb}(z_{\rm id})<t_{\rm cross}=2\,
  r_{200}/V_{200}$. These are systems too dynamically young to be in
  virial equilibrium. The bottom curves in each panel show systems
  whose formation times suggest that they should have
  higher-than-average concentrations; i.e., those with
  $0.25<t_{\rm h}/t_{\rm cross}<0.75$ (see shaded area in
  Fig.~\ref{Figczh}). Different colours are used to indicate results
  from MS-I, MS-II, and MS-XXL, as indicated in the legend.}
\label{FigMFrel}
\end{figure*}

\subsection{Concentration estimates and virialization}
\label{SecConcVir}

Massive halos formed hierarchically assemble late and they are
therefore expected to have accreted a large fraction of their mass in
a short period of time, mainly in the form of major mergers. As a
merger proceeds, oscillations in the mass profile drive fluctuations
in the concentration as the system relaxes to equilibrium. 

We illustrate this in Fig.~\ref{FigCVir}, where we show the evolution
of a dark matter halo that undergoes a late-time merger. The system
chosen for this illustration is halo Aq-F from the Aquarius Project
\citep{Springel2008b}. The top panel of Fig.~\ref{FigCVir} shows the
evolution of the virial mass of the most massive progenitor,
normalized to the mass at $z=0$, as well as that of $r_{200}/r_{\rm
  h}$. The bottom two panels monitor the dynamical relaxation criteria
introduced above; when at least one of them fails (i.e., strides
outside the shaded areas) the halo would be deemed unrelaxed. We use
larger symbols to indicate the times along the evolution when this
happens.


Fig.~\ref{FigCVir} shows that Aq-F undergoes a major merger with a
massive secondary halo (``subhalo'', for short) at $z \sim
0.4$. Different evolutionary stages are easily identified: the merger
begins at $z_{\rm acc}=0.43$, when the subhalo first enters the
virial radius of Aq-F; it reaches pericenter shortly thereafter at
$z_{\rm per}=0.35$. The subhalo is largely disrupted then, and the
remnant halo bounces once more before quickly virializing
afterward. 

The effect of these oscillations on the concentration are shown by the
solid squares that track the evolution of $r_{200}/r_{\rm h}$ in the
top panel of Fig.~\ref{FigCVir}. Concentrations first drop at $t_{\rm
  acc}$, reach a maximum at $t_{\rm per}$, and decrease again as the
subhalo reaches its second apocenter before settling down. Since first
accretion, it takes at least one crossing time (defined as $t_{\rm
  cross}=2\, r_{200}/V_{200}$) for the system to virialize. The arrow
labelled $t_{\rm rel}$ indicates the moment when one crossing time has
elapsed since accretion.  

For a system like Aq-F that nearly doubles its
virial mass in a merger, the oscillations have large amplitudes:
roughly a factor of $\sim 2$ in $r_{200}/r_{\rm h}$, which translates
into {\it a fluctuation of almost one order of magnitude in
  concentration}.  Fig.~\ref{FigCVir} also shows that the dynamical
relaxation criteria introduced above can be effective at identifying
systems undergoing these transitions. Unless such criteria are
introduced, the average concentration of a halo sample that includes a
large fraction of recently-assembled systems may differ
  significantly from that of their equilibrium counterparts.

\subsection{Concentration and formation redshift}
\label{Secczh}

Guided by the results discussed in the previous subsection, we explore
in Fig.~\ref{Figczh} the dependence of concentration on formation
redshift. Because halos of different mass collapse on average at
different times, we choose halos in the narrow mass range, $3.49 <
\log (M_{200}/10^{10} h^{-1} M_{\odot}) < 3.79$, indicated by the
shaded vertical band in Fig.~\ref{FigMc}. This is close to the
characteristic clustering mass at $z=0$ ($1.35<\nu<1.55$), but
at $z=3$ corresponds to the rarest, most massive systems present in
the MS-XXL simulation ($4.12<\nu<4.73$). 

The three panels of Fig.~\ref{Figczh} show results for halos
identified at $z_{\rm id}=0$, $1$, and $3$. The bottom panels show the
half-mass radius concentration estimate, $r_{200}/r_{\rm h}$, as a
function of the half-mass formation redshift, $z_{\rm h}$, defined as
the time when the most massive progenitor first reaches one half of
the virial mass of the halo at $z_{\rm id}$. Tickmarks on the right
indicate the corresponding NFW-based concentrations. Connected symbols
trace the median concentration; thin lines indicate the 25th and 75th
percentiles.

The bottom panels of Fig.~\ref{Figczh} show that, as expected, halos
that form earlier tend to be more concentrated. However, this holds
{\it only} for halos that are ``old'' enough at the time of
identification to have had a chance to relax to equilibrium: the
concentration of halos formed less than one crossing-time ago (i.e., $t_{\rm
  h}=t_{\rm lb}(z_{\rm h})-t_{\rm lb}(z_{\rm id})<t_{\rm cross}$)
depends non-monotonically on $z_{\rm h}$. (We use $t_{\rm lb}$ to
denote cosmic lookback times since $z=0$.) The median $r_{200}/r_{\rm h}$
trend reverts and has a pronounced maximum when $t_{\rm h} \sim t_{\rm
  cross}/2$. The same behaviour is observed at all three redshifts,
and reflects the oscillations in the structure of halos that have
recently accreted a large fraction of their mass as they relax to
equilibrium.  Halos formed in the period $0.25<t_{\rm h}/t_{\rm
  cross}<0.75$ (highlighted by the shaded area in Fig.~\ref{Figczh})
are caught at $z_{\rm id}$ during a transient state of maximal
concentration.

In general, the more massive the halo the larger the fraction of
systems that have assembled recently and that are therefore out of
equilibrium. This is shown in the top panels of Fig.~\ref{Figczh},
where the histograms indicate the distribution of formation redshifts
of all clusters (in blue) and of relaxed clusters (in
orange). Clearly, very few halos that have assembled less than one
crossing time ago pass the relaxation criteria. Indeed, at $z=3$, when
halos in this mass range are very rare the majority are less than one
crossing time ``old'' and fewer than $3\%$ of them are deemed to be in
equilibrium.  In particular, note that at $z=3$ a large fraction of
systems are at the maximally-compressed stage that corresponds to the
first pericentric passage of the accreted material, thus increasing
the average concentration. 

This suggests that the aforementioned upturn in concentration at high
masses is due to the increasing prevalence of systems caught during
this particular evolutionary stage. We show this explicitly in
Fig.~\ref{FigMFrel}, where we plot, as a function of virial mass, the
fraction of halos with lookback formation times satisfying $t_{\rm
  h}<t_{\rm cross}$ (solid lines) and $0.25<t_{\rm h}/t_{\rm
  cross}<0.75$ (dashed lines). The former characterizes halos formed
too recently to have reached equilibrium; the latter indicates the
fraction likely found in a temporary, highly-concentrated state.
Although recently-assembled systems of any mass are susceptible to
this effect, the increasing fraction of systems in the latter stage
with halo mass (or $\nu$) provides clear support for our
interpretation of the upturn as driven by systems passing through a
short-lived, highly-concentrated stage of their evolution.

Finally, Fig.~\ref{FigMFrel} makes clear that the fraction of clusters
out of equilibrium is not just a function of $\nu$, but also of
redshift. For example, at $z_{\rm id}=0$, about $12\%$ of $\nu \sim
2.5$ systems were formed less than one crossing time ago. At fixed
$\nu$, this fraction increases to $42\%$ and $69\%$ at $z_{\rm id}=1$
and $z_{\rm id}=3$, respectively. This increase in the fraction of
``unrelaxed'' systems at given $\nu$ explains why \citet{Prada2011}
find that the median concentration, at given $\nu$, increases with
redshift (see their Fig.7). The origin of this effect may be traced to
the mass-dependent shape of the power spectrum. At fixed $\nu$, the
higher the redshift the lower the halo mass and the shallower the
slope of the mass fluctuation spectrum, $d\log\sigma/d\log
M$. Shallower spectra imply faster structure growth
\citep{Efstathiou1988}, implying that, for a given $\nu$, clusters at
high-redshift are being assembled comparably ``faster'' than their
lower-redshift counterparts. This increases the out-of-equilibrium
fraction with redshift and pushes the median concentration even higher.

\section{Summary and Conclusions}
\label{SecConc}

We have analyzed dark matter halos identified at different redshifts
in the Millennium Simulation series (MS-I, MS-II, and MS-XXL) to
investigate the origin of the recently-reported upturn in the
mass-concentration relation of rare, massive halos. We estimate
concentrations from the ratio of virial-to-half mass radius,
$r_{200}/r_{\rm h}$, assuming that halos follow an NFW profile. Our
analysis confirms that, as a function of the dimensionless mass
parameter $\nu(M,z)=\delta_{\rm crit}(z)/\sigma(M,z)$, the median
concentration declines gradually until it reaches a minimum
at $\nu\sim 3$, then increases toward higher values of $\nu$. This upturn,
however, is not present when only dynamically-relaxed halos are
retained for analysis, a clear indication that out-of-equilibrium
effects drive the non-monotonic behaviour of the mass-concentration
relation.

Further inspection demonstrates that the upturn is due to systems
caught in a state of ``maximal compression'' coincident with the first
pericentric passage of recently-accreted material. This affects
primarily halos that have just accreted a large fraction of their
mass, and especially those where the lookback time to formation
coincides approximately with the time it takes to bring material from
the virial radius of the halo to the center, roughly
$r_{200}/V_{200}$.  The fraction of systems in this transient stage
increases systematically with halo mass, causing the upturn in the
$M(c)$ relation that becomes manifest at extremely high masses, where
maximally-compressed halos dominate. 


The non-monotonic nature of the mass-concentration relation
illustrates the fact that, due to their recent assembly, {\it the vast
  majority of very massive galaxy clusters are generally expected to
  be out of dynamical equilibrium}. It further warns about the
applicability of simple power-law scalings for various observational
mass proxies, such as velocity dispersion, X-ray
luminosity/temperature, or SZ decrement, especially when applied to
the rarest, most massive clusters. Any sample of such systems is
likely to be dominated by systems currently out-of equilibrium,
implying that deviations from simple power-laws are to be generally
expected. These deviations lead to substantial, correlated
  scatter between different mass estimators that should scale in non-trivial
  ways with cluster mass and redshift.  Our findings suggest that the
dynamical state of galaxy clusters must be carefully and explicitly
taken into account before cluster studies may be used as reliable
tools of precision cosmology.

\bsp
\label{lastpage}


\begin{center}
\begin{table}
  \caption{Halo sample considered in our analysis. N$_{\rm halos}$
    lists the total number of halos with $N_{200}\geq 5000$ in each simulation;
    N$_{\rm rel}$ the total number of ``relaxed'' halos (see Section~\ref{SecAnal}); and $f_{\rm rel}$ is the relaxed halo fraction.}
\begin{tabular}{c c r r c c c c}\hline \hline
Run  & $z_{\rm id}$ & N$_{\rm halo}$ & N$_{\rm rel}$ & $f_{\rm rel}$ &\\\hline
MSII &  0       & 55,152         & 42,395         &  0.803        &\\ 
     &  1       & 64,650         & 35,359         &  0.547        &\\ 
     &  3       & 48,321         & 10,412         &  0.215        &\\\hline

MSI  &  0       & 83,196         & 57,782         &  0.695        &\\ 
     &  1       & 64,907         & 22,560         &  0.348        &\\ 
     &  3       & 6,560          & 402            &  0.061        &\\\hline

XXL  &  0       & 2,101,739      & 598,603        &  0.285        &\\ 
     &  1       & 910,043        & 102,475        &  0.113        &\\ 
     &  3       & 4,799          & 112            &  0.023        &\\\hline \hline
\end{tabular}
\label{TabHaloSample}
\end{table}
\end{center}

\section*{acknowledgements}
We would like to thank Simon White, Volker Springel and Gerard Lemson for
useful discussion, and the Virgo Consortium for access to the MS data. 
ADL acknowledges finacial support from the SFB (956) from the Deutsche 
Forschungsgemeinschaft. REA is supported by Advanced Grant 246797  “GALFORMOD” 
from the European Research Council. MB-K acknowledges support from the Southern 
California Center for Galaxy Evolution, a multi-campus research program funded 
by the University of California Office of Research.

\bibliographystyle{mn2e}
\bibliography{paper}

\begin{thebibliography}{}

\bibitem[\protect\citeauthoryear{{Allen}, {Evrard} \& {Mantz}}{{Allen}
  et~al.}{2011}]{Allen2011}
{Allen} S.~W.,  {Evrard} A.~E.,    {Mantz} A.~B.,  2011, \araa, 49, 409

\bibitem[\protect\citeauthoryear{{Angulo}, {Springel}, {White}, {Jenkins},
  {Baugh} \& {Frenk}}{{Angulo} et~al.}{2012}]{Angulo2012}
{Angulo} R.~E.,  {Springel} V.,  {White} S.~D.~M.,  {Jenkins} A.,  {Baugh}
  C.~M.,    {Frenk} C.~S.,  2012, ArXiv e-prints: 1203.3216

\bibitem[\protect\citeauthoryear{{Boylan-Kolchin}, {Springel}, {White},
  {Jenkins} \& {Lemson}}{{Boylan-Kolchin} et~al.}{2009}]{Boylan-Kolchin2009}
{Boylan-Kolchin} M.,  {Springel} V.,  {White} S.~D.~M.,  {Jenkins} A.,
  {Lemson} G.,  2009, \mnras, 398, 1150

\bibitem[\protect\citeauthoryear{{Bullock}, {Kolatt}, {Sigad}, {Somerville},
  {Kravtsov}, {Klypin}, {Primack} \& {Dekel}}{{Bullock}
  et~al.}{2001}]{Bullock2001}
{Bullock} J.~S.,  {Kolatt} T.~S.,  {Sigad} Y.,  {Somerville} R.~S.,  {Kravtsov}
  A.~V.,  {Klypin} A.~A.,  {Primack} J.~R.,    {Dekel} A.,  2001, \mnras, 321,
  559

\bibitem[\protect\citeauthoryear{{Cunha}, {Huterer} \& {Frieman}}{{Cunha}
  et~al.}{2009}]{Cunha2009}
{Cunha} C.,  {Huterer} D.,    {Frieman} J.~A.,  2009, \prd, 80, 063532

\bibitem[\protect\citeauthoryear{{Davis}, {Efstathiou}, {Frenk} \&
  {White}}{{Davis} et~al.}{1985}]{Davis1985}
{Davis} M.,  {Efstathiou} G.,  {Frenk} C.~S.,    {White} S.~D.~M.,  1985, \apj,
  292, 371

\bibitem[\protect\citeauthoryear{{Efstathiou}, {Frenk}, {White} \&
  {Davis}}{{Efstathiou} et~al.}{1988}]{Efstathiou1988}
{Efstathiou} G.,  {Frenk} C.~S.,  {White} S.~D.~M.,    {Davis} M.,  1988,
  \mnras, 235, 715

\bibitem[\protect\citeauthoryear{{Eke}, {Navarro} \& {Steinmetz}}{{Eke}
  et~al.}{2001}]{Eke2001}
{Eke} V.~R.,  {Navarro} J.~F.,    {Steinmetz} M.,  2001, \apj, 554, 114

\bibitem[\protect\citeauthoryear{{Evrard} \& {Henry}}{{Evrard} \&
  {Henry}}{1991}]{Evrard1991}
{Evrard} A.~E.,  {Henry} J.~P.,  1991, \apj, 383, 95

\bibitem[\protect\citeauthoryear{{Henry}, {Evrard}, {Hoekstra}, {Babul} \&
  {Mahdavi}}{{Henry} et~al.}{2009}]{Henry2009}
{Henry} J.~P.,  {Evrard} A.~E.,  {Hoekstra} H.,  {Babul} A.,    {Mahdavi} A.,
  2009, \apj, 691, 1307

\bibitem[\protect\citeauthoryear{{Kaiser}}{{Kaiser}}{1986}]{Kaiser1986}
{Kaiser} N.,  1986, \mnras, 222, 323

\bibitem[\protect\citeauthoryear{{Kaiser}}{{Kaiser}}{1991}]{Kaiser1991}
{Kaiser} N.,  1991, \apj, 383, 104

\bibitem[\protect\citeauthoryear{{Klypin}, {Trujillo-Gomez} \&
  {Primack}}{{Klypin} et~al.}{2011}]{Klypin2011}
{Klypin} A.~A.,  {Trujillo-Gomez} S.,    {Primack} J.,  2011, \apj, 740, 102

\bibitem[\protect\citeauthoryear{{Mantz}, {Allen}, {Ebeling} \&
  {Rapetti}}{{Mantz} et~al.}{2008}]{Mantz2008}
{Mantz} A.,  {Allen} S.~W.,  {Ebeling} H.,    {Rapetti} D.,  2008, \mnras, 387,
  1179

\bibitem[\protect\citeauthoryear{{Navarro}, {Frenk} \& {White}}{{Navarro}
  et~al.}{1995}]{Navarro1995}
{Navarro} J.~F.,  {Frenk} C.~S.,    {White} S.~D.~M.,  1995, \mnras, 275, 720

\bibitem[\protect\citeauthoryear{{Navarro}, {Frenk} \& {White}}{{Navarro}
  et~al.}{1996}]{Navarro1996}
{Navarro} J.~F.,  {Frenk} C.~S.,    {White} S.~D.~M.,  1996, \apj, 462, 563

\bibitem[\protect\citeauthoryear{{Navarro}, {Frenk} \& {White}}{{Navarro}
  et~al.}{1997}]{Navarro1997}
{Navarro} J.~F.,  {Frenk} C.~S.,    {White} S.~D.~M.,  1997, \apj, 490, 493

\bibitem[\protect\citeauthoryear{{Neto}, {Gao}, {Bett}, {Cole}, {Navarro},
  {Frenk}, {White}, {Springel} \& {Jenkins}}{{Neto} et~al.}{2007}]{Neto2007}
{Neto} A.~F.,  {Gao} L.,  {Bett} P.,  {Cole} S.,  {Navarro} J.~F.,  {Frenk}
  C.~S.,  {White} S.~D.~M.,  {Springel} V.,    {Jenkins} A.,  2007, \mnras,
  381, 1450

\bibitem[\protect\citeauthoryear{{Planck Collaboration}, {Aghanim}, {Arnaud},
  {Ashdown}, {Aumont}, {Baccigalupi}, {Balbi}, {Banday}, {Barreiro},
  {Bartelmann} \& et al.}{{Planck Collaboration} et~al.}{2011}]{Planck2011c}
{Planck Collaboration} {Aghanim} N.,  {Arnaud} M.,  {Ashdown} M.,  {Aumont} J.,
   {Baccigalupi} C.,  {Balbi} A.,  {Banday} A.~J.,  {Barreiro} R.~B.,
  {Bartelmann} M.,    et al. 2011, \aap, 536, A12

\bibitem[\protect\citeauthoryear{{Prada}, {Klypin}, {Cuesta}, {Betancort-Rijo}
  \& {Primack}}{{Prada} et~al.}{2011}]{Prada2011}
{Prada} F.,  {Klypin} A.~A.,  {Cuesta} A.~J.,  {Betancort-Rijo} J.~E.,
  {Primack} J.,  2011, ArXiv e-prints: 1104.5130

\bibitem[\protect\citeauthoryear{{Rozo}, {Rykoff}, {Evrard}, {Becker}, {McKay},
  {Wechsler}, {Koester}, {Hao}, {Hansen}, {Sheldon}, {Johnston}, {Annis} \&
  {Frieman}}{{Rozo} et~al.}{2009}]{Rozo2009}
{Rozo} E.,  {Rykoff} E.~S.,  {Evrard} A.,  {Becker} M.,  {McKay} T.,
  {Wechsler} R.~H.,  {Koester} B.~P.,  {Hao} J.,  {Hansen} S.,  {Sheldon} E.,
  {Johnston} D.,  {Annis} J.,    {Frieman} J.,  2009, \apj, 699, 768

\bibitem[\protect\citeauthoryear{{Springel}, {Wang}, {Vogelsberger}, {Ludlow},
  {Jenkins}, {Helmi}, {Navarro}, {Frenk} \& {White}}{{Springel}
  et~al.}{2008}]{Springel2008b}
{Springel} V.,  {Wang} J.,  {Vogelsberger} M.,  {Ludlow} A.,  {Jenkins} A.,
  {Helmi} A.,  {Navarro} J.~F.,  {Frenk} C.~S.,    {White} S.~D.~M.,  2008,
  \mnras, 391, 1685

\bibitem[\protect\citeauthoryear{{Springel}, {White}, {Jenkins}, {Frenk},
  {Yoshida}, {Gao}, {Navarro}, {Thacker}, {Croton}, {Helly}, {Peacock}, {Cole},
  {Thomas}, {Couchman}, {Evrard}, {Colberg} \& {Pearce}}{{Springel}
  et~al.}{2005}]{Springel2005a}
{Springel} V.,  {White} S.~D.~M.,  {Jenkins} A.,  {Frenk} C.~S.,  {Yoshida} N.,
   {Gao} L.,  {Navarro} J.,  {Thacker} R.,  {Croton} D.,  {Helly} J.,
  {Peacock} J.~A.,  {Cole} S.,  {Thomas} P.,  {Couchman} H.,  {Evrard} A.,
  {Colberg} J.,    {Pearce} F.,  2005, \nat, 435, 629

\bibitem[\protect\citeauthoryear{{Springel}, {White}, {Tormen} \&
  {Kauffmann}}{{Springel} et~al.}{2001}]{Springel2001b}
{Springel} V.,  {White} S.~D.~M.,  {Tormen} G.,    {Kauffmann} G.,  2001,
  \mnras, 328, 726

\bibitem[\protect\citeauthoryear{{Vikhlinin}, {Kravtsov}, {Burenin}, {Ebeling},
  {Forman}, {Hornstrup}, {Jones}, {Murray}, {Nagai}, {Quintana} \&
  {Voevodkin}}{{Vikhlinin} et~al.}{2009}]{Vikhlinin2009}
{Vikhlinin} A.,  {Kravtsov} A.~V.,  {Burenin} R.~A.,  {Ebeling} H.,  {Forman}
  W.~R.,  {Hornstrup} A.,  {Jones} C.,  {Murray} S.~S.,  {Nagai} D.,
  {Quintana} H.,    {Voevodkin} A.,  2009, \apj, 692, 1060

\bibitem[\protect\citeauthoryear{{Wechsler}, {Bullock}, {Primack}, {Kravtsov}
  \& {Dekel}}{{Wechsler} et~al.}{2002}]{Wechsler2002}
{Wechsler} R.~H.,  {Bullock} J.~S.,  {Primack} J.~R.,  {Kravtsov} A.~V.,
  {Dekel} A.,  2002, \apj, 568, 52

\bibitem[\protect\citeauthoryear{{Zhao}, {Jing}, {Mo} \& {B{\"o}rner}}{{Zhao}
  et~al.}{2003}]{Zhao2003a}
{Zhao} D.~H.,  {Jing} Y.~P.,  {Mo} H.~J.,    {B{\"o}rner} G.,  2003, \apjl,
  597, L9

\end{thebibliography}

\end{document}